\documentclass{article}
\usepackage{graphicx}
\usepackage{amsmath}

\newtheorem{theorem}{Theorem}

\newtheorem{lemma}[theorem]{Lemma}

\newtheorem{proposition}[theorem]{Proposition}

\newenvironment{proof}[1][Proof]{\textbf{#1.} }{\ \rule{0.5em}{0.5em}}
\input{tcilatex}

\begin{document}

\title{On the Extreme Flights of One-Sided L\'{e}vy Processes\thanks{%
This manuscript is an extended version of a contribution to a special \emph{%
Physica A} volume in honor of Shlomo Havlin on his sixtieth birthday.}}
\author{Iddo Eliazar \\
Recanaty Faculty of Management\\
Tel Aviv University\\
Tel Aviv 69978\\
Israel\\
\emph{eliazar@post.tau.ac.il}\\
\& \and Joseph Klafter \\
School of Chemistry\\
Sackler Faculty of Exact Sciences\\
Tel Aviv University\\
Tel Aviv 69978\\
Israel\\
\emph{klafter@post.tau.ac.il}}
\maketitle

\begin{abstract}
We explore the statistical behavior of the order statistics of the flights
of One-sided L\'{e}vy Processes (OLPs). We begin with the study of the
extreme flights of general OLPs, and then focus on the class of selfsimilar
processes, investigating the following issues: (i) the inner hierarchy of
the extreme flights - for example: how big is the $7^{\text{th}}$ largest
flight relative to the $2^{\text{nd}}$ largest one?; and, (ii) the relative
contribution of the extreme flights to the entire `flight aggregate' - for
example: how big is the $3^{\text{rd}}$ largest flight relative to the OLP's
value?. Furthermore, we show that all `hierarchical' results obtained - but
not the `aggregate' results - are explicitly extendable to the class of OLPs
with arbitrary power-law flight tails (which is far larger than the
selfsimilar class).

\bigskip

\end{abstract}

\section{Introduction}

\qquad L\'{e}vy processes - random motions with stationary and independent
increments - constitute one of the most important and fundamental family of
stochastic processes. Special examples of the L\'{e}vy family include
Brownian motion (Wiener process), Poisson processes, and Compound Poisson
processes. Since their introduction in the 1930s, by the French
mathematician Paul L\'{e}vy \cite{L25}-\cite{L65}, L\'{e}vy processes were
studied extensively by both theoreticians and applied scientists. The
literature on L\'{e}vy processes is vast, and their range of applications
encompasses numerous fields of science and engineering. See \cite{GK}-\cite
{BMR} for the theory of L\'{e}vy processes, and \cite{UZ}-\cite{Hav} and
references therein for their applications.

Amongst the family of L\'{e}vy processes, the class of One-sided L\'{e}vy
Processes (OLPs), also referred to as L\'{e}vy Subordinators, is of special
importance. OLPs satisfy the additional requirement of non-negativity of
their increments, rendering them monotone non-decreasing and non-negative
valued. OLPs are natural models for random flow of positive valued
quantities. Examples include: mass, energy, and time in physical systems;
work and costumers in queueing systems; data in communication systems;
claims in insurance; etc.

Unlike Brownian motion - whose sample paths are continuous, the sample paths
of OLPs - regardless of their statistics - are always purely \emph{%
discontinuous}. The propagation of OLPs is conducted only by \emph{flights}
(jumps), and not by any sort of continuous motion. This is due to the fact
that OLPs are continuum superpositions, or \emph{aggregates}, of Poisson
processes.

Since the structure of OLPs is that of a `flight aggregate', questions of
the following type arise naturally: how big is the \emph{largest flight}?
how big is the $n^{\text{th}}$ largest flight? how big is the $m^{\text{th}}$
largest flight \emph{relative} to the $n^{\text{th}}$ largest flight ($m>n$%
)? how big is the largest flight relative to the \emph{entire} aggregate?
how big is the \emph{combined} contribution of the $n^{\text{th}}$ largest
flights relative to the entire aggregate? - that is, questions regarding the 
\emph{order statistics} of the flights of OLPs.

The study of the order statistics of series of Independent and Identically
Distributed (IID) random variables is a well established field of
probability called \emph{Extreme Value Theory} (EVT). This theory
originated, in the 1920s and 30s, with the pioneering works of von
Bortkiewicz \cite{Bor}, Fr\'{e}chet \cite{Fre}, Fisher \& Tippett \cite{FT},
von Mises \cite{Mis}, Weibull, and Gumbel \cite{Gum} (the theory's three
possible types of limiting distributions are named after Gumbel (``type
1''), Fr\'{e}chet (``type 2''), and Weibull (``type 3'')). A rigorous
theoretical framework was presented in 1943 by Gnedenko \cite{Gne}. The
statistical analysis of extreme values is of major importance in the
analysis of rare and `catastrophic' events such as floods in hydrology,
large claims in insurance, crashes in finance, material failure in corrosion
analysis, etc. See \cite{EKM}-\cite{TR} and references therein for both the
theory and applications of modern EVT. See also \cite{MANA} for a recent
application to the study of complex networks.

OLPs are the continuous time counterparts of non-negative IID random
sequences (viewed as discrete time stochastic processes). When passing from
discrete time to continuous time the question ``what was the largest
observation amongst the first $N$ observations?'' transforms to ``what was
the largest flight that occurred up to time $T$?''. More generally, the
analogue of the \emph{order statistics} of a discrete sequence (largest
observation, second largest observation, etc) is the order statistics of the 
\emph{flights} of a continuous time process (largest flight, second largest
flight, etc). Hence, the study of the extreme flights of OLPs is also a
natural sequel to the EVT of IID random variables.

Furthermore, the only possible non-trivial \emph{scaling limits} of
non-negative IID random sequences are the, so called, \emph{selfsimilar }%
OLPs. This special class of motions, which occupies a predominant role in
both the theory and applications of stochastic processes, is composed of all
OLPs which are invariant under changes of scale. Since EVT is, in essence,
an \emph{asymptotic} theory (providing limiting results as the total number
of observations $N$ tends to infinity) we shall devote our main emphasis in
this work to the exploration of this special class of OLPs.

\bigskip

The paper is organized as follows; We begin, in section \ref{2}, with an
investigation of the extreme flights of general OLPs (following a short
review of these processes). In sections \ref{3} and \ref{4} we focus on the
class of selfsimilar OLPs. Section \ref{3} studies of the \emph{internal
hierarchy} of the extreme flights - dealing with questions of the type ``how
big are the extreme flights relative to each other?''. Section \ref{4}, on
the other hand, studies the magnitude of the extreme flights relative to the
entire process - the `\emph{flight aggregate}'. Furthermore, we explain why
(and how) the results of section \ref{3} are extendable to the entire class
of OLPs with power-law flight tails, and, on the other hand, why the results
of section \ref{4} do \emph{not} extend.

Throughout the manuscript: \textbf{P}$\left( \cdot \right) $ = Probability;
and, \textbf{E}$\left[ \cdot \right] $ = Expectation.

\section{\label{2}OLPs and their extreme flights}

In this section we give a short review of OLPs, establish a few preliminary
results regarding the distribution of their extreme flights, and conclude
with the special class of selfsimilar OLPs and its connection to the
Fr\'{e}chet (``type 2'') extreme value distribution \cite{KN}.

\subsection{L\'{e}vy processes and OLPs}

A stochastic process is said to be \emph{L\'{e}vy} if it is continuous in
probability and has stationary and independent increments. A \emph{OLP} is a
L\'{e}vy process with non-negative increments.

\bigskip

\textbf{Characterization of OLPs}

The celebrated L\'{e}vy-Khinchin formula (see, for example, \cite{ST} or 
\cite{Ber}) asserts that L\'{e}vy processes can be decomposed into two
independent parts: (i) a purely continuous part, which is a Brownian motion;
and, (ii) a purely discontinuous part, which is a superposition of Poisson
processes. Since Brownian motion is symmetric, OLPs can not include a
continuous part and are hence purely discontinuous (pure jump) processes.

A OLP $L=\left( L(t)\right) _{t\geq 0}$ is characterized by its \emph{%
Laplace transform} (its `spectral representation' in Laplace space): 
\begin{equation}
\text{\textbf{E}}\left[ \exp \left\{ -\omega L(t)\right\} \right] =\exp
\left\{ -\Psi (\omega )\cdot t\right\} \text{ \ \ ; \ \ }\omega \geq 0\text{
.}  \label{2.00}
\end{equation}
The function $\Psi $ is called the \emph{L\'{e}vy characteristic} of the OLP 
$L$ (in the literature, $\Psi $ is also referred to as the spectral
characteristic, or symbol, of $L$).

\bigskip

\textbf{The L\'{e}vy measure}

If $L$ is a Poisson process with flights of size $x_{0}$ ($x_{0}>0$) and
rate $\lambda _{0}$, then its L\'{e}vy characteristic is $\Psi (\omega
)=\left( 1-\exp \left\{ -\omega x_{0}\right\} \right) \lambda _{0}$. If $L$
is a superposition of $N$ independent Poisson processes - process $n$, $%
n=1,2,\cdots ,N$, having flights of size $x_{n}$ ($x_{n}>0$)\ and rate $%
\lambda _{n}$ - then its L\'{e}vy characteristic is given by 
\begin{equation}
\Psi (\omega )=\sum_{n=1}^{N}\left( 1-\exp \left\{ -\omega x_{n}\right\}
\right) \lambda _{n}\text{ .}  \label{2.01}
\end{equation}
Hence, passing from (\ref{2.01}) to a continuum limit, where flights of size 
$x$ ($x>0$) occur at rate $\lambda (dx)$, we arrive at 
\begin{equation}
\Psi (\omega )=\int_{0}^{\infty }\left( 1-\exp \left\{ -\omega x\right\}
\right) \lambda (dx)\text{ .}  \label{2.02}
\end{equation}

Equation (\ref{2.02}) is the L\'{e}vy-Khinchin representations (in Laplace
space) for OLPs. The rate $\lambda (\cdot )$ - the L\'{e}vy measure of $L$ -
is a measure on the non-negative half line $(0,\infty )$ satisfying the
integrability condition $\int_{0}^{\infty }\min \{x,1\}\lambda (dx)$. The
total rate $\int_{0}^{\infty }\lambda (dx)$ could certainly be \emph{infinite%
} - not due to non-integrability at $x=\infty $ but, rather, due to possible
non-integrability at $x=0$ (intuitively, large flights can occur only
rarely, but tiny flights may occur very frequently). A OLP is Compound
Poisson if and only if its L\'{e}vy measure has finite total mass.

Given a L\'{e}vy measure $\lambda (\cdot )$ it is natural to introduce its
`cumulative distribution function'. However, since $\lambda (\cdot )$ might
have infinite total mass (due to possible non-integrability\ at $x=0$) we
need to define the `cumulative distribution function' by integrating
(backwards) from $x=\infty $ rather than (forward) from $x=0$: 
\begin{equation}
\Lambda (x)=\int_{x}^{\infty }\lambda (dy)\text{ \ \ ; \ \ }x>0\text{ .}
\label{2.03}
\end{equation}
We henceforth refer to $\Lambda $ as the OLP's \emph{flight tail}. The
meaning of the flight tail is straightforward: $\Lambda (x)$ is the \emph{%
rate} at which flights of size $>x$ occur.

\bigskip

\textbf{Examples}

We give a few examples of classes of OLPs. The first two have finite
L\'{e}vy measure, whereas the other have infinite L\'{e}vy measure;

\begin{enumerate}
\item  Compound Poisson with exponentially-distributed jumps ($a>0$): 
\begin{equation*}
\lambda (dx)=a\exp \{-ax\}dx\text{ \ \ \ \ }\left( \Lambda (x)=\exp
\{-ax\}\right) \text{ ,}
\end{equation*}
\begin{equation*}
\Psi (\omega )=\frac{\omega }{a+\omega }\text{ .}
\end{equation*}

\item  Compound Poisson with Gamma-distributed jumps ($a,p>0$): 
\begin{equation*}
\lambda (dx)=\exp \{-ax\}x^{p-1}dx\text{ ,}
\end{equation*}
\begin{equation*}
\Psi (\omega )=\Gamma (p)\left( \frac{1}{a^{p}}-\frac{1}{(a+\omega )^{p}}%
\right) \text{ .}
\end{equation*}

\item  Gamma processes - OLPs with Gamma-distributed increments ($a>0$): 
\begin{equation*}
\lambda (dx)=\frac{\exp \{-ax\}}{x}dx\text{ ,}
\end{equation*}
\begin{equation*}
\Psi (\omega )=\ln \left( 1+\frac{\omega }{a}\right) \text{ .}
\end{equation*}

\item  Selfsimilar (`fractal') OLPs ($0<\alpha <1$): 
\begin{equation*}
\lambda (dx)=\frac{\alpha }{x^{1+\alpha }}dx\text{ \ \ \ \ }\left( \Lambda
(x)=\frac{1}{x^{\alpha }}\right) \text{ ,}
\end{equation*}
\begin{equation*}
\Psi (\omega )=\Gamma (1-\alpha )\omega ^{\alpha }\text{ .}
\end{equation*}
\end{enumerate}

\subsection{The extreme flights of OLPs}

Given a OLP $L=\left( L(t)\right) _{t\geq 0}$ with L\'{e}vy measure $\lambda
(\cdot )$ (flight tail $\Lambda (\cdot )$), we set 
\begin{equation}
X_{1}(t)>X_{2}(t)>X_{3}(t)>\cdots  \label{2.10}
\end{equation}
to be the \emph{order sequence} of the \emph{flights} of the OLP $L$, i.e; $%
X_{n}(t)$ denotes the $n^{\text{th}}$\emph{\ largest flight} that occurred
during the time interval $[0,t]$. If the OLP has finite L\'{e}vy measure $%
\Lambda (0)<\infty $ (i.e; if it is Compound Poisson) then, up to time $t$,
there will have occurred only a finite number of flights (Poisson with rate $%
t\Lambda (0)$, to be exact) and the order sequence will hence be finite. On
the other hand, if the OLP has infinite L\'{e}vy measure $\Lambda (0)=\infty 
$ then the order sequence will be infinite.

Since the Poisson distribution will appear time and again in the sequel, we
introduce the following shorthand notation for its probability frequencies: 
\begin{equation}
P_{n}(\mu )=\frac{\mu ^{n}}{n!}\exp (-\mu )\text{ .}  \label{2.11}
\end{equation}
That is; $P_{n}(\mu )$ denotes the probability that a Poisson random
variable with rate $\mu $ ($\mu >0$) equals the integer value $n$ ($%
n=0,1,2,\cdots $).

The \emph{joint} distribution of the flights of a OLP is of major
importance. Given a time $t$ ($t>0$) and an interval $I\subset (0,\infty )$,
let $\Pi (t;I)$ denote the number of flights of $L$, during the time period $%
[0,t]$, whose size laid in the interval $I$. Due to the
`Poissonian-superposition' structure of $L$ the random variable $\Pi (t;I)$
is Poisson-distributed with rate $t\int_{I}\lambda \left( dx\right) $.
Moreover, if $I_{1},\cdots ,I_{K}$ are \emph{disjoint} intervals then $\Pi
(t;I_{1}),\cdots ,\Pi (t;I_{K})$ are \emph{independent} random variables and
hence 
\begin{equation}
\text{\textbf{P}}\left( \Pi (t;I_{k})=n_{k}\text{ ; }k=1,\cdots ,K\right)
=\prod_{k=1}^{K}P_{n_{k}}\left( t\int_{I_{k}}\lambda \left( dx\right)
\right) \text{ ,}  \label{2.12}
\end{equation}
$\forall n_{1},\cdots ,n_{K}\in \{0,1,2,\cdots \}$. For further details see 
\cite{Ber}.

\bigskip

\textbf{The greatest flight}

The distribution of the maximal flight $X_{1}(t)$ is given by: 
\begin{equation}
\text{\textbf{P}}\left( X_{1}(t)\leq x\right) =\exp \{-t\Lambda (x)\}\text{ .%
}  \label{2.13}
\end{equation}
The explanation of (\ref{2.13}) is given by the following straightforward
deduction: 
\begin{equation*}
\left. 
\begin{array}{l}
\text{\textbf{P}}\left( X_{1}(t)\leq x\right) \\ 
\\ 
=\text{\textbf{P}}\left( \Pi (t;(x,\infty ))=0\right) \\ 
\\ 
=P_{0}\left( t\int_{x}^{\infty }\lambda \left( ds\right) \right) \\ 
\\ 
=\exp \{-t\Lambda (x)\}\text{ .}
\end{array}
\right.
\end{equation*}

Equation (\ref{2.13}) enables the `shock tolerance' design of systems. We
explain;

Assume that $L=\left( L(t)\right) _{t\geq 0}$ is the inflow to a system,
which cannot absorb flights (i.e; inflow surges) of size greater than a
tolerance level $l$. The following natural engineering question arises: what
should the tolerance level $l$ be so that to ensure that the system would
withstand the time period $[0,T]$ with no failure, with probability greater
than $1-\delta $ ($0<\delta <1$)? The answer, deduced by a simple inversion
of (\ref{2.13}) is 
\begin{equation*}
l=\Lambda ^{-1}\left( \frac{1}{T}\ln \left( \frac{1}{1-\delta }\right)
\right) \text{ .}
\end{equation*}

The maximal flight, viewed as a stochastic \emph{process} $\left(
X_{1}(t)\right) _{t\geq 0}$, is \emph{Markovian}. Indeed, when at level $x$ (%
$x>0$): \textbf{(i)} the process has to wait an exponential time, with rate $%
\Lambda (x)$, till transition; and then, \textbf{(ii) }the process will
transit to the level $y$ ($y>x$) with probability $\lambda \left( dy\right)
/\Lambda (x)$. In other words, the infinitesimal generator of the process $%
\left( X_{1}(t)\right) _{t\geq 0}$ is given by the integral operator: 
\begin{equation*}
\left( \mathcal{L}_{1}\varphi \right) (x)=\int_{x}^{\infty }\left( \varphi
(y)-\varphi (x)\right) \lambda \left( dy\right) \text{ .}
\end{equation*}

\bigskip

\textbf{The }$n^{\text{th}}$\textbf{\ runner-up}

The distribution of the $n^{\text{th}}$ largest-order flight $X_{n}(t)$ is
given by: 
\begin{equation}
\text{\textbf{P}}\left( X_{n}(t)\leq x\right) =\sum_{k=0}^{n-1}P_{k}\left(
t\Lambda (x)\right) \text{ .}  \label{2.14}
\end{equation}
The explanation of (\ref{2.14}) is analogous to the deduction of (\ref{2.13}%
): 
\begin{equation*}
\left. 
\begin{array}{l}
\text{\textbf{P}}\left( X_{n}(t)\leq x\right) \\ 
\\ 
=\text{\textbf{P}}\left( \Pi (t;(x,\infty ))\leq n-1\right) \\ 
\\ 
=\sum_{k=0}^{n-1}P_{k}\left( t\int_{x}^{\infty }\lambda \left( ds\right)
\right) \\ 
\\ 
=\sum_{k=0}^{n-1}P_{k}\left( t\Lambda (x)\right) \text{ .}
\end{array}
\right.
\end{equation*}

\bigskip

\textbf{The order sequence}

The entire order sequence can be constructed/simulated sequentially,
according to the iterative scheme ($y<x$): 
\begin{equation}
\text{\textbf{P}}\left( X_{n+1}(t)\leq y\text{ \TEXTsymbol{\vert} }%
X_{n}(t)=x\right) =\exp \{-t\left( \Lambda (y)-\Lambda (x)\right) \}\text{\ ,%
}  \label{2.15}
\end{equation}
together with the initial condition (\ref{2.13}). Indeed, $\forall y<x$ we
have 
\begin{equation*}
\left. 
\begin{array}{l}
\text{\textbf{P}}\left( X_{n+1}(t)\leq y\text{ \TEXTsymbol{\vert} }%
X_{n}(t)=x\right) \\ 
\\ 
=\text{\textbf{P}}\left( \Pi (t;(y,x))=0\right) \\ 
\\ 
=P_{0}\left( t\int_{y}^{x}\lambda \left( ds\right) \right) \\ 
\\ 
=\exp \{-t\left( \Lambda (y)-\Lambda (x)\right) \}\text{ .}
\end{array}
\right.
\end{equation*}

Furthermore, the joint distribution of the truncated order sequence $\left(
X_{1}(t),\cdots ,X_{K}(t)\right) $ is given, for a decreasing sequence $%
\infty >b_{1}>a_{1}>b_{2}>a_{2}>\cdots >b_{K}>a_{K}=0$, by the following
formula: 
\begin{equation}
\left. 
\begin{array}{c}
\text{\textbf{P}}\left( a_{k}<X_{k}(t)<b_{k}\text{ ; }k=1,\cdots ,K\right)
\\ 
= \\ 
\exp \{-t\Lambda (b_{K})\}\cdot t^{K-1}\prod_{k=1}^{K-1}\left( \Lambda
(a_{k})-\Lambda (b_{k})\right)
\end{array}
\right. .  \label{2.16}
\end{equation}
The derivation of (\ref{2.16}) is analogous to the derivation of (\ref{2.13}%
)-(\ref{2.15}).

\subsection{\label{3Fer}Self-similarity and the Fr\'{e}chet distribution}

In the proceeding sections we shall focus on selfsimilar OLPs, i.e; OLPs
with flight tail $\Lambda (x)=1/x^{\alpha }$ ($0<\alpha <1$). For these
processes the distribution of the maximal flight $X_{1}(t)$ is given by (due
to (\ref{2.13})): 
\begin{equation}
\text{\textbf{P}}\left( X_{1}(t)\leq x\right) =\exp \{-t/x^{\alpha }\}\text{
.}  \label{2.17}
\end{equation}

\bigskip

\textbf{The Fr\'{e}chet distribution}

The probability law (\ref{2.17}) is known as the Fr\'{e}chet distribution,
or as the ``type 2'' extreme value distribution \cite{KN}. It emerges as the
asymptotic probability law of the maximum of Independent and Identically
Distributed (IID) heavy tailed random variables. We explain;

Let $Y_{1},\cdots ,Y_{N}$ be a sequence of IID, non-negative, random
variables with heavy (`fat') tails: \textbf{P}$\left( Y>y\right) \sim
a/y^{\alpha }$ as $y\rightarrow \infty $ (for some $a,\alpha >0$). Then, the
appropriately scaled sequence maximum converges, in law, to the Fr\'{e}chet
distribution: 
\begin{equation}
\text{\textbf{P}}\left( \frac{1}{N^{1/\alpha }}\max \{Y_{1},\cdots
,Y_{N}\}\leq y\right) \underset{N\rightarrow \infty }{\longrightarrow }\exp
\{-a/y^{\alpha }\}\text{ .}  \label{2.18}
\end{equation}
Indeed; 
\begin{equation*}
\left. 
\begin{array}{l}
\text{\textbf{P}}\left( \frac{1}{N^{1/\alpha }}\max \{Y_{1},\cdots
,Y_{N}\}\leq y\right) \\ 
\\ 
=\left( 1-\text{\textbf{P}}\left( Y>N^{1/\alpha }y\right) \right) ^{N} \\ 
\\ 
\sim \left( 1-\frac{a/y^{\alpha }}{N}\right) ^{N} \\ 
\\ 
\sim \exp \{-a/y^{\alpha }\}\text{ .}
\end{array}
\right.
\end{equation*}

Note, however, the discrepancy between (\ref{2.17}) and (\ref{2.18}): in (%
\ref{2.17}) we are restricted to the parameter range\footnote{%
Selfsimilar OSLMs with exponent $\alpha \geq 1$ are trivial: $L(t)\equiv
const\cdot t$ if $\alpha =1$, and $L(t)\equiv 0$ if $\alpha >1$.} $0<\alpha
<1$, whereas in (\ref{2.18}) the exponent $\alpha $ may admit any positive
value. This discrepancy stems from the following reason;

Consider the IID sequence $Y_{1},\cdots ,Y_{N}$ introduced above. The
scaling limit of its maximum is given by (\ref{2.18}), for all $\alpha >0$.
However, in the scaling limit of their sum (aggregate) $S_{N}=Y_{1}+\cdots
+Y_{N}$ a phase transition occurs when passing from $\alpha <1$ to $\alpha
>1 $: \textbf{(i)} for $\alpha <1$ the appropriate scaling is $%
S_{N}/N^{1/\alpha }$ and the limiting distribution is a selfsimilar L\'{e}vy
law of order $\alpha $; but, \textbf{(ii)} for $\alpha >1$ the scaling is
the universal Law of Large Numbers scaling - $S_{N}/N$ - leading to the
deterministic limit \textbf{E}$\left[ Y\right] $. Hence, scaling limits of
non-negative IID sums - yielding the selfsimilar OLPs - are restricted to
the `L\'{e}vy range' $0<\alpha <1$, whereas scaling limits of non-negative
IID maxima admit the entire `Fr\'{e}chet range' $\alpha >0$.

This gap between the `L\'{e}vy range' and the `Fr\'{e}chet range' can be
bridged by taking scaling limits of OLPs with power-law flight tails;

\bigskip

\textbf{Scaling limits}

Let $L$ be a OLPs whose flight tail asymptotics (as $x\rightarrow \infty )$
are $\Lambda (x)\sim 1/x^{\alpha }$, for some exponent $\alpha >0$.
Introduce the \emph{scaled} order sequence, $%
X_{1}^{(c)}(t)>X_{2}^{(c)}(t)>X_{3}^{(c)}(t)>\cdots $, given by 
\begin{equation*}
X_{n}^{(c)}(t)=\frac{1}{c}X_{n}(c^{\alpha }t)
\end{equation*}
(where $c>0$ is a scaling factor), and consider the \emph{scaling limit }of
the order sequence, i.e; the limit, in law, as $c\rightarrow \infty $, of
the scaled sequence: $X_{1}^{(\infty )}(t)>X_{2}^{(\infty
)}(t)>X_{3}^{(\infty )}(t)>\cdots $. For the scaling limit $X_{1}^{(\infty
)}(t)$ we have: 
\begin{equation*}
\text{\textbf{P}}\left( X_{1}^{(\infty )}(t)\leq x\right) =\exp
\{-t/x^{\alpha }\}\text{ ,}
\end{equation*}
and now the exponent $\alpha $ is \emph{not }(!) restricted to the `L\'{e}vy
range' $0<\alpha <1$ - as it is in (\ref{2.17}).

In fact, we obtain the following counterpart of (\ref{2.16}): 
\begin{equation}
\left. 
\begin{array}{c}
\text{\textbf{P}}\left( a_{k}<X_{k}^{(\infty )}(t)<b_{k}\text{ ; }k=1,\cdots
,K\right) \\ 
= \\ 
\exp \left\{ \frac{-t}{(b_{K})^{\alpha }}\right\} \cdot
t^{K-1}\prod_{k=1}^{K-1}\left( \frac{1}{(a_{k})^{\alpha }}-\frac{1}{%
(b_{k})^{\alpha }}\right)
\end{array}
\right. ,  \label{2.19}
\end{equation}
which holds for the entire `Fr\'{e}chet range' $\alpha >0$. The deduction of
(\ref{2.19}) stems from (\ref{2.16}) combined with the scaling procedure
described above. Indeed, for $c>>1$ we have: 
\begin{equation*}
\left. 
\begin{array}{l}
\text{\textbf{P}}\left( a_{k}<X_{k}^{(\infty )}(t)<b_{k}\text{ ; }k=1,\cdots
,K\right) \\ 
\\ 
=\text{\textbf{P}}\left( ca_{k}<X_{n}(c^{\alpha }t)<cb_{k}\text{ ; }%
k=1,\cdots ,K\right) \\ 
\\ 
=\exp \{-(c^{\alpha }t)\Lambda (cb_{K})\}\cdot (c^{\alpha
}t)^{K-1}\prod_{k=1}^{K-1}\left( \Lambda (ca_{k})-\Lambda (cb_{k})\right) \\ 
\\ 
\sim \exp \left\{ \frac{-t}{(b_{K})^{\alpha }}\right\} \cdot
t^{K-1}\prod_{k=1}^{K-1}\left( \frac{1}{(a_{k})^{\alpha }}-\frac{1}{%
(b_{k})^{\alpha }}\right) \text{ .}
\end{array}
\right.
\end{equation*}

\section{\label{3}The hierarchy of extreme flights}

In this section we explore the \emph{hierarchical} structure of the order
sequence of selfsimilar OLPs. We assume, throughout this section, that the
OLP $L$ is $\alpha $-selfsimilar, i.e; that its flight tail is $\Lambda
(x)=1/x^{\alpha }$.

The \emph{Beta distribution} will emerge naturally and play a key role in
this section \cite{Fel}. We denote by $B_{n,m}(x)$, $x\in \lbrack 0,1]$, the
cumulative distribution function of a $Beta(n,m)$ distribution ($n,m>0$): 
\begin{equation}
B_{n,m}(x)=\int_{0}^{x}\frac{\Gamma (n+m)}{\Gamma (n)\Gamma (m)}%
u^{n-1}(1-u)^{m-1}du\text{ .}  \label{3.00}
\end{equation}

\bigskip

Let us begin with the investigation of the pairwise hierarchy;

\subsection{Pairwise hierarchy}

In this subsection we study the statistics of the ratio $X_{n+m}(t)/X_{n}(t)$%
, i.e; the size of the $(n+m)^{\text{th}}$ largest-order flight \emph{%
relative} to the size of the $n^{\text{th}}$ largest-order flight. Clearly,
this ratio admits values ranging in the unit interval. We assert that the
cumulative distribution function of this ratio is independent of time $t$
and is given,$\forall u\in \lbrack 0,1]$, by: 
\begin{equation}
\text{\textbf{P}}\left( \frac{X_{n+m}(t)}{X_{n}(t)}\leq u\right)
=B_{n,m}(u^{\alpha })\text{ .}  \label{3.10}
\end{equation}
Equation (\ref{3.10}) is a particular case of proposition \ref{3b} which
will be presented in the following subsection.

Moreover, the moments of the ratio $X_{n+m}(t)/X_{n}(t)$ are derived from (%
\ref{3.10}); $\forall p>0$ we have: 
\begin{equation}
\text{\textbf{E}}\left[ \left( \frac{X_{n+m}(t)}{X_{n}(t)}\right) ^{p}\right]
=\prod_{k=n}^{n+m-1}\frac{k}{k+p/\alpha }\text{ .}  \label{3.11}
\end{equation}
The proof of (\ref{3.11}) is brought in the appendix.

We point out two special cases:

\bigskip

`\textbf{Consecutive relativity'}

The statistics of the ratio $X_{n+1}(t)/X_{n}(t)$ are given by: 
\begin{equation*}
\text{\textbf{P}}\left( \frac{X_{n+1}(t)}{X_{n}(t)}\leq u\right) =u^{\alpha
n}\text{ ,}
\end{equation*}
and 
\begin{equation*}
\text{\textbf{E}}\left[ \left( \frac{X_{n+1}(t)}{X_{n}(t)}\right) ^{p}\right]
=\frac{n}{n+p/\alpha }\text{ .}
\end{equation*}

\bigskip

`\textbf{Maximal relativity'}

The statistics of the ratio $X_{1+m}(t)/X_{1}(t)$ are given by: 
\begin{equation*}
\text{\textbf{P}}\left( \frac{X_{1+m}(t)}{X_{1}(t)}\leq u\right) =1-\left(
1-u^{\alpha }\right) ^{m}\text{ ,}
\end{equation*}
and 
\begin{equation*}
\text{\textbf{E}}\left[ \left( \frac{X_{1+m}(t)}{X_{1}(t)}\right) ^{p}\right]
=\prod_{k=1}^{m}\frac{k}{k+p/\alpha }\text{ .}
\end{equation*}

\subsection{Multidimensional hierarchy}

In the previous subsection we investigated the \emph{pairwise} hierarchy of
the extreme flights. Now we turn to study the \emph{multidimensional}
hierarchical structure of the order sequence: given an increasing sequence
of integers $n_{1}<\cdots <n_{K}<n_{K+1}$ we wish to compute the joint
distribution of the vector of ratios 
\begin{equation}
\left( \frac{X_{n_{2}}(t)}{X_{n_{1}}(t)},\frac{X_{n_{3}}(t)}{X_{n_{2}}(t)}%
,\cdots ,\frac{X_{n_{K+1}}(t)}{X_{n_{K}}(t)}\right) \text{ .}  \label{3.20}
\end{equation}
To that end we have;

\begin{proposition}
\label{3b}The multidimensional cumulative distribution function of the
vector of ratios (\ref{3.20}) is independent of time $t$ and is given, $%
\forall u_{1},\cdots ,u_{K}\in \lbrack 0,1]$, by 
\begin{equation}
\text{\textbf{P}}\left( \frac{X_{n_{k+1}}(t)}{X_{n_{k}}(t)}\leq u_{k}\text{
; }k=1,\cdots ,K\right)
=\prod_{k=1}^{K}B_{n_{k},n_{k+1}-n_{k}}((u_{k})^{\alpha })\text{ .}
\label{3.21}
\end{equation}
\end{proposition}

The proof of proposition \ref{3b} is brought in the appendix. This
proposition can be re-stated as follows:

\smallskip

\begin{center}
\emph{The ratios in (\ref{3.20}) are independent and are distributed
according to (\ref{3.10}).}
\end{center}

\bigskip

In particular, proposition \ref{3b} implies that: 
\begin{equation}
\text{\textbf{P}}\left( \frac{X_{k+1}(t)}{X_{k}(t)}\leq u_{k}\text{ ; }%
k=1,\cdots ,K\right) =\prod_{k=1}^{K}(u_{k})^{\alpha k}\text{ ,}
\label{3.22}
\end{equation}
$\forall u_{1},\cdots ,u_{K}\in \lbrack 0,1]$.

\bigskip

\textbf{Multidimensional `maximal relativity'}

The distribution of the truncated order sequence relative to the \emph{%
maximal flight} 
\begin{equation}
\left( \frac{X_{2}(t)}{X_{1}(t)},\frac{X_{3}(t)}{X_{1}(t)},\cdots ,\frac{%
X_{K+1}(t)}{X_{1}(t)}\right) \text{ ,}  \label{3.23}
\end{equation}
rather than its multidimensional `consecutive relativity' counterpart (\ref
{3.22}), is given, for a decreasing sequence $1\geq a_{1}>\cdots
>a_{K}>a_{K+1}=0$, by the formula: 
\begin{equation}
\left. 
\begin{array}{c}
\text{\textbf{P}}\left( a_{k+1}<\frac{X_{k+1}(t)}{X_{1}(t)}<a_{k}\text{ ; }%
k=1,\cdots ,K\right) \\ 
= \\ 
\Gamma (K)(a_{K})^{\alpha K}\prod_{k=1}^{K-1}\left( \frac{1}{%
(a_{k+1})^{\alpha }}-\frac{1}{(a_{k})^{\alpha }}\right)
\end{array}
\right. .  \label{3.24}
\end{equation}
The proof of (\ref{3.24})\ is brought in the appendix.

\bigskip

\textbf{A `Fr\'{e}chet remark'}

All the results of this section may be extended from the restricted
selfsimilar `L\'{e}vy range' $0<\alpha <1$ to the entire `Fr\'{e}chet range' 
$\alpha >0$ following the scaling procedure described in subsection \ref
{3Fer}: replace the order sequence $X_{1}(t)>X_{2}(t)>X_{3}(t)>\cdots $ of
an $\alpha $-selfsimilar OLP ($0<\alpha <1$) by the scaling limit $%
X_{1}^{(\infty )}(t)>X_{2}^{(\infty )}(t)>X_{3}^{(\infty )}(t)>\cdots $ of
the order sequence of a OLP whose flight tail asymptotics are $\Lambda
(x)\sim 1/x^{\alpha }$ ($\alpha >0$).

\section{\label{4}The extreme flights \emph{vs} the aggregate}

In the previous section we studied the behavior of the extreme flights
relative to \emph{each other}. We now turn to study the extreme flights
relative to the \emph{entire }process $L$. Since $L$ is a (continuum)
superposition of Poisson processes, it is, in fact, a `\emph{flight aggregate%
}' (where flights of size $x$ occur with rate $\lambda (dx)$). Hence, we are
interested in investigating the relative contribution of the largest
flights, up to time $t$, to the entire `fight aggregate' $L(t)$. As in the
previous section, we assume in this section that the OLP $L$ is $\alpha $%
-selfsimilar.

\bigskip

Before we begin, let us introduce the family of functions $\left\{ G_{\alpha
}\right\} _{0<\alpha <1}$, defined on the non-negative half line ($\omega
\geq 0$), and given by the integral formula 
\begin{equation}
G_{\alpha }(\omega )=\alpha \int_{0}^{1}\frac{1-\exp \{-\omega u\}}{%
u^{1+\alpha }}du\text{ ,}  \label{4.00}
\end{equation}
or, equivalently, by the power series 
\begin{equation}
G_{\alpha }(\omega )=-\alpha \sum_{m=1}^{\infty }\frac{(-1)^{m}}{m-\alpha }%
\cdot \frac{\omega ^{m}}{m!}\text{ .}  \label{4.01}
\end{equation}

\subsection{The greatest flight \emph{vs} the aggregate}

Let $\xi (t)$ denote the magnitude of the ``entire flight aggregate, up to
time $t$, \emph{except} for the maximal flight'', relative to the maximal
flight: 
\begin{equation*}
\xi (t)=\frac{L(t)-X_{1}(t)}{X_{1}(t)}\text{ .}
\end{equation*}
We assert that the ratio $\xi (t)$ is independent of time $t$ and that its
Laplace transform is given by: 
\begin{equation}
\text{\textbf{E}}\left[ \exp \{-\omega \xi (t)\}\right] =\frac{1}{%
1+G_{\alpha }(\omega )}\text{ .}  \label{4.10}
\end{equation}
Equation (\ref{4.10}) is a particular case of proposition \ref{4c} which
will be presented in the following subsection. Let us denote by $\mathcal{G}%
_{\alpha }$ the probability law whose Laplace transform is given by the
right hand side of (\ref{4.10}).

\bigskip

Equation (\ref{4.10}) implies that the mean and variance of $\xi (t)$ are
independent of time $t$, and are given, respectively, by 
\begin{equation}
\text{\textbf{E}}\left[ \xi (t)\right] =\frac{\alpha }{1-\alpha }\text{ ,}
\label{4.11}
\end{equation}
and 
\begin{equation}
\text{\textbf{Var}}\left( \xi (t)\right) =\frac{\alpha }{(1-\alpha
)^{2}(2-\alpha )}\text{ .}  \label{4.12}
\end{equation}
The derivation of (\ref{4.11})-(\ref{4.12}) is obtained by differentiating (%
\ref{4.10}) and using the power expansion (\ref{4.01}). Combining (\ref{4.11}%
)-(\ref{4.12}) together we also have 
\begin{equation}
\text{\textbf{Var}}\left( \frac{\xi (t)}{\text{\textbf{E}}\left[ \xi (t)%
\right] }\right) =\frac{1}{\alpha (2-\alpha )}\text{ .}  \label{4.13}
\end{equation}
From (\ref{4.11})-(\ref{4.13}) we see that:

When $\alpha \rightarrow 1$ the underlying OLP $L$ converges to the
degenerate deterministic linear limit $\equiv t$, and hence all flights tend
to zero. Indeed, from (\ref{4.11}) we obtain that \textbf{E}$\left[ \xi (t)%
\right] \rightarrow \infty $ as $\alpha \rightarrow 1$. That is, the size of
the aggregate, relative to the greatest flight, tends to infinity. Moreover,
the variance of normalized ratio $\xi (t)/$\textbf{E}$\left[ \xi (t)\right] $
converges, as $\alpha \rightarrow 1$, to $1$.

On the other hand, when $\alpha \rightarrow 0$ the tails of the OLP $L$
become infinitely `fat' and hence the greatest flight should dominate. And
indeed, we obtain that \textbf{E}$\left[ \xi (t)\right] \rightarrow 0$ as $%
\alpha \rightarrow 0$. That is, the size of the aggregate, relative to the
greatest flight, tends to zero. In fact, the dominance of the largest flight
is so great that even \textbf{Var}$\left( \xi (t)\right) \rightarrow 0$ as $%
\alpha \rightarrow 0$.

The value $\alpha =1/2$ turns out to be the ``break-even'' point where the
mean of ratio $\xi (t)$ equals $1$. That is, the entire aggregate splits, on
average, half-half between the largest flight and ``all the rest''. As $%
\alpha $ `moves up' towards $1$ the weight of ``all the rest'' takes the
lead and eventually dominates, and vice-versa as $\alpha $ `moves down'
towards $0$.

\bigskip

Another immediate consequences of (\ref{4.10}), after noting that $%
X_{1}(t)/L(t)=1/(1+\xi (t))$, is that the cumulative distribution function
of the ratio $X_{1}(t)/L(t)$ is independent of time $t$ and is given, $%
\forall u\in \lbrack 0,1]$, by 
\begin{equation}
\text{\textbf{P}}\left( \frac{X_{1}(t)}{L(t)}\leq u\right) =\text{\textbf{P}}%
\left( \xi \geq \frac{1}{u}-1\right) \text{ ,}  \label{4.14}
\end{equation}
where $\xi $ is $\mathcal{G}_{\alpha }$-distributed.

\subsection{The top-$n$ \emph{vs} the aggregate}

We now proceed to study the contribution of the `top-$n$' flights, $%
X_{1}(t),X_{2}(t),\cdots ,X_{n}(t)$, relative to the entire aggregate $L(t)$%
. The key result is:

\begin{proposition}
\label{4c}The random vector 
\begin{equation}
\left( \frac{L(t)-\left( X_{1}(t)+\cdots +X_{n+1}(t)\right) }{X_{n+1}(t)}%
\text{ ; }\frac{X_{1}(t)}{X_{n+1}(t)},\cdots ,\frac{X_{n}(t)}{X_{n+1}(t)}%
\right)  \label{4.21}
\end{equation}
is independent of time $t$ and equal, in law, to the vector 
\begin{equation*}
(\xi _{1}+\cdots +\xi _{n+1}\text{ ; }Y_{(1)},\cdots ,Y_{(n)})
\end{equation*}
where: (i) $\xi _{1},\cdots ,\xi _{n+1},Y_{1},\cdots ,Y_{n}$ are independent
random variables; (ii) $\xi _{1},\cdots ,\xi _{n+1}$\ are $\mathcal{G}%
_{\alpha }$-distributed; and, (iii) $Y_{(1)},\cdots ,Y_{(n)}$ are the order
statistics of $Y_{1},\cdots ,Y_{n}$ which, in turn, are Pareto$(\alpha )$%
-distributed: \textbf{P}$\left( Y>y\right) =1/y^{\alpha }$, $y\geq 1$.
\end{proposition}

The proof of proposition \ref{4c} is brought in the appendix.

\bigskip

Noting that $Y_{(1)}+\cdots +Y_{(n)}=Y_{1}+\cdots +Y_{n}$, proposition \ref
{4c} implies the following pair of corollaries:

\textbf{(i)} The ratio $X_{n+1}(t)/L(t)$ - the contribution of the $(n+1)^{%
\text{th}}$ largest flight, relative to the entire aggregate - is equal, in
law, to 
\begin{equation*}
\begin{array}{c}
\left( 1+\xi _{1}+\cdots +\xi _{n+1}+Y_{1}+\cdots +Y_{n}\right) ^{-1}
\end{array}
\text{.}
\end{equation*}
Hence, $\forall u\in \lbrack 0,1]$ we have: 
\begin{equation}
\text{\textbf{P}}\left( \frac{X_{n+1}(t)}{L(t)}\leq u\right) =\text{\textbf{P%
}}\left( \xi _{1}+\cdots +\xi _{n+1}+Y_{1}+\cdots +Y_{n}\geq \frac{1}{u}%
-1\right) \text{ .}  \label{4.22}
\end{equation}

\textbf{(ii)} The ratio $\left( X_{1}(t)+\cdots +X_{n}(t)\right) /L(t)$ -
the combined contribution of the $n$ largest flights, relative to the entire
aggregate - is equal, in law, to: 
\begin{equation*}
\left( 1+\frac{1+\xi _{1}+\cdots +\xi _{n+1}}{Y_{1}+\cdots +Y_{n}}\right)
^{-1}\text{ .}
\end{equation*}
Hence, $\forall u\in \lbrack 0,1]$ we have: 
\begin{equation}
\text{\textbf{P}}\left( \frac{X_{1}(t)+\cdots +X_{n}(t)}{L(t)}\leq u\right) =%
\text{\textbf{P}}\left( \frac{1+\xi _{1}+\cdots +\xi _{n+1}}{Y_{1}+\cdots
+Y_{n}}\geq \frac{1}{u}-1\right) \text{ .}  \label{4.23}
\end{equation}

\bigskip

\textbf{A Fr\'{e}chet remark}

The results of this section are \emph{not} extendable from the restricted
selfsimilar `L\'{e}vy range' $0<\alpha <1$ to the entire `Fr\'{e}chet range' 
$\alpha >0$. This is since the \emph{scaling procedure} described in
subsection \ref{3Fer} - which holds for any $\alpha >0$ when regarding the 
\emph{extremes }- fails to hold for the \emph{aggregate} when $\alpha >1$
(since, as explained in subsection \ref{3Fer}, the limiting behavior of the
aggregate undergoes a phase transition when passing from the `L\'{e}vy
kingdom' $\alpha <1$ to the `Law of Large Numbers kingdom' $\alpha >1$).

\section{\label{6}Appendix: proofs}

\subsection{\label{6.1}Equation (\ref{3.11}): moments of the pairwise
hierarchy}

\begin{proof}
Combining (\ref{3.00}) and (\ref{3.10}), the probability density function of
the ratio $X_{n+m}(t)/X_{n}(t)$ is given by 
\begin{equation*}
f_{n,m}(u)=\frac{\Gamma (n+m)}{\Gamma (n)\Gamma (m)}(u^{\alpha
})^{n-1}(1-u^{\alpha })^{m-1}\cdot \alpha u^{\alpha -1}\text{ ,}
\end{equation*}
Hence, changing the integration variable to $x=u^{\alpha }$ and using Beta
integrals, we obtain 
\begin{equation*}
\left. 
\begin{array}{l}
\text{\textbf{E}}\left[ \left( \frac{X_{n+m}(t)}{X_{n}(t)}\right) ^{p}\right]
\\ 
\\ 
=\int_{0}^{1}u^{p}\cdot f_{n,m}(u)du \\ 
\\ 
=\frac{\Gamma (n+m)}{\Gamma (n)\Gamma (m)}\int_{0}^{1}x^{(n+p/\alpha
)-1}(1-x)^{m-1}dx \\ 
\\ 
=\frac{\Gamma (n+m)}{\Gamma (n)}\frac{\Gamma (n+p/\alpha )}{\Gamma
(n+m+p/\alpha )} \\ 
\\ 
=\prod_{k=n}^{n+m-1}\frac{k}{k+p/\alpha }\text{ .}
\end{array}
\right.
\end{equation*}
\end{proof}

\subsection{\label{6.2}proposition \ref{3b}}

We begin with a lemma;

\begin{lemma}
\label{6a}Let $\Lambda (x)=1/x^{\alpha }$, $x>0$, and $n,m>0$. Then, $%
\forall x>0$ and $\forall 0<u<1$ we have 
\begin{equation}
\frac{1}{\Gamma (n)\Gamma (m)}\int_{x/u}^{\infty }\Lambda (y)^{n-1}\left(
\Lambda (x)-\Lambda (y)\right) ^{m-1}\lambda (dy)=\frac{\Lambda (x)^{n+m-1}}{%
\Gamma (n+m)}B_{n,m}(u^{\alpha })\text{ .}  \label{6.20}
\end{equation}
\end{lemma}

\begin{proof}
First, note that we can re-write the left hand side of (\ref{6.20}) as
follows 
\begin{equation}
\frac{\Lambda (x)^{n+m-1}}{\Gamma (n)\Gamma (m)}\cdot \int_{x/u}^{\infty
}\left( \frac{\Lambda (y)}{\Lambda (x)}\right) ^{n-1}\left( 1-\frac{\Lambda
(y)}{\Lambda (x)}\right) ^{m-1}\frac{\lambda (dy)}{\Lambda (x)}\text{ .}
\label{6.21}
\end{equation}
Now, using the change of variable $s=\Lambda (y)/\Lambda (x)$, the integral
part of (\ref{6.21}) equals 
\begin{equation}
\int_{0}^{u^{\alpha }}s^{n-1}\left( 1-s\right) ^{m-1}ds=\frac{\Gamma
(n)\Gamma (m)}{\Gamma (n+m)}B_{n,m}(u^{\alpha })\text{ .}  \label{6.22}
\end{equation}
Substituting (\ref{6.22}) back into (\ref{6.21}) concludes the proof.
\end{proof}

\bigskip

We are ready now to prove the proposition;

\bigskip

\begin{proof}
First, we use (\ref{2.12}) in order to compute the multidimensional density
function of the vector $\left( X_{n_{1}}(t),\cdots
,X_{n_{K}}(t),X_{n_{K+1}}(t)\right) $: 
\begin{equation*}
\left. 
\begin{array}{l}
\text{\textbf{P}}\left( X_{n_{k}}(t)\in dx_{k}\text{ ; }k=1,\cdots
,K+1\right) \\ 
\\ 
=\prod_{k=1}^{K+1}P_{1}\left( t\lambda (dx_{k})\right)
P_{n_{k}-n_{k-1}-1}\left( t\int_{x_{k}}^{x_{k-1}}\lambda \left( ds\right)
\right) \\ 
\\ 
=\prod_{k=1}^{K+1}\left( t\lambda (dx_{k})\right) \left( \frac{\left[
t\left( \Lambda (x_{k})-\Lambda (x_{k-1})\right) \right] ^{n_{k}-n_{k-1}-1}}{%
\Gamma \left( n_{k}-n_{k-1}\right) }\exp \left\{ -t\left( \Lambda
(x_{k})-\Lambda (x_{k-1})\right) \right\} \right) \\ 
\\ 
=t^{n_{K+1}}\exp \left\{ -t\Lambda (x_{K+1})\right\} \cdot \prod_{k=1}^{K+1}%
\frac{\left( \Lambda (x_{k})-\Lambda (x_{k-1})\right) ^{n_{k}-n_{k-1}-1}}{%
\Gamma \left( n_{k}-n_{k-1}\right) }\lambda (dx_{k}) \\ 
\\ 
=f\left( x_{1},\cdots ,x_{K},x_{K+1}\right) \cdot \lambda (dx_{1})\cdot
\cdots \cdot \lambda (dx_{K})\cdot \lambda (dx_{K+1})\text{ .}
\end{array}
\right.
\end{equation*}
$\forall $decreasing sequence $\infty =x_{0}>x_{1}>\cdots >x_{K}>x_{K+1}>0$,
and where $n_{0}=0$.

\bigskip

Hence 
\begin{equation}
\left. 
\begin{array}{c}
\text{\textbf{P}}\left( \frac{X_{n_{k+1}}(t)}{X_{n_{k}}(t)}\leq u_{k}\text{
; }k=1,\cdots ,K\right) \\ 
= \\ 
\\ 
\int_{x_{K+1}}^{\infty }\lambda (dx_{K+1})\int_{x_{K}=x_{K+1}/u_{K}}^{\infty
}\lambda (dx_{K})\cdots \\ 
\\ 
\cdots \int_{x_{2}=x_{3}/u_{2}}^{\infty }\lambda
(dx_{2})\int_{x_{1}=x_{2}/u_{1}}^{\infty }\lambda (dx_{1})f\left(
x_{1},\cdots ,x_{K},x_{K+1}\right) \text{ .}
\end{array}
\right.  \label{6.23}
\end{equation}
Now, computing the multiple integral (\ref{6.23}), using lemma \ref{6a}
repeatedly, yields 
\begin{equation*}
\text{\textbf{P}}\left( \frac{X_{n_{k+1}}(t)}{X_{n_{k}}(t)}\leq u_{k}\text{
; }k=1,\cdots ,K\right) =I\cdot
\prod_{k=1}^{K}B_{n_{k},n_{k+1}-n_{k}}((u_{k})^{\alpha })\text{ ,}
\end{equation*}
where 
\begin{equation*}
I=\int_{0}^{\infty }\frac{t^{n_{K+1}}}{\Gamma \left( n_{K+1}\right) }\exp
\left\{ -t\Lambda (x_{K+1})\right\} \Lambda (x_{K+1})^{n_{K+1}-1}\lambda
(dx_{K+1})\text{ ,}
\end{equation*}
which, in turn, must equal $1$.
\end{proof}

\subsection{\label{6.3}Equation (\ref{3.24}): multidimensional `maximal
relativity'}

\begin{proof}
(Recall that $\Lambda (x)=1/x^{\alpha }$)

Analogously to the proof of proposition \ref{3b}, we have 
\begin{equation}
\left. 
\begin{array}{c}
\text{\textbf{P}}\left( a_{k+1}<\frac{X_{k+1}(t)}{X_{k}(t)}<a_{k}\text{ ; }%
k=1,\cdots ,K\right) \\ 
= \\ 
\\ 
\int_{0}^{\infty }\lambda (dx_{1})\int_{a_{2}x_{1}}^{a_{1}x_{1}}\lambda
(dx_{2})\cdots \int_{a_{K}x_{1}}^{a_{K-1}x_{1}}\lambda
(dx_{K})\int_{0}^{a_{K}x_{1}}\lambda (dx_{K+1})f\left( x_{1},\cdots
,x_{K},x_{K+1}\right) \text{ ,}
\end{array}
\right.  \label{6.30}
\end{equation}
where 
\begin{equation*}
f\left( x_{1},\cdots ,x_{K},x_{K+1}\right) =t^{K+1}\exp \left\{ -t\Lambda
(x_{K+1})\right\} \text{ .}
\end{equation*}

Now, $\forall k=2,\cdots ,K$ we have 
\begin{equation}
\left. 
\begin{array}{l}
\int_{a_{k}x_{1}}^{a_{k-1}x_{1}}\lambda (dx_{k}) \\ 
\\ 
=\Lambda (a_{k}x_{1})-\Lambda (a_{k-1}x_{1}) \\ 
\\ 
=\left( \Lambda (a_{k})-\Lambda (a_{k-1})\right) \Lambda (x_{1})\text{ .}
\end{array}
\right.  \label{6.31}
\end{equation}
And, using the change of variable $s=\Lambda (x_{K+1})$, 
\begin{equation}
\left. 
\begin{array}{l}
\int_{0}^{a_{K}x_{1}}\lambda (dx_{K+1})t\exp \left\{ -t\Lambda
(x_{K+1})\right\} \\ 
\\ 
=\int_{\Lambda (a_{K})\Lambda (x_{1})}^{\infty }t\exp \{-ts\}ds \\ 
\\ 
=\exp \{-t\Lambda (a_{K})\Lambda (x_{1})\}\text{ .}
\end{array}
\right.  \label{6.32}
\end{equation}

Substituting (\ref{6.31}) and (\ref{6.32}) into the multiple integral on the
right hand side of (\ref{6.30}) yields 
\begin{equation}
I\cdot \prod_{k=2}^{K}\left( \Lambda (a_{k})-\Lambda (a_{k-1})\right) \text{
,}  \label{6.33}
\end{equation}
where 
\begin{equation*}
I=t^{K}\int_{0}^{\infty }\exp \{-t\Lambda (a_{K})\Lambda (x_{1})\}\Lambda
(x_{1})^{K-1}\lambda (dx_{1})\text{ .}
\end{equation*}
However, using the change of variable $s=\Lambda (x_{1})$ gives 
\begin{equation}
I=t^{K}\int_{0}^{\infty }\exp \{-t\Lambda (a_{K})s\}s^{K-1}ds=\frac{\Gamma
(K)}{\Lambda (a_{K})^{K}}  \label{6.34}
\end{equation}

Finally, combining (\ref{6.33}) and (\ref{6.34}) together implies that the
right hand side of (\ref{6.30}) equals 
\begin{equation*}
\frac{\Gamma (K)}{\Lambda (a_{K})^{K}}\prod_{k=2}^{K}\left( \Lambda
(a_{k})-\Lambda (a_{k-1})\right) =\Gamma (K)(a_{K})^{\alpha
K}\prod_{k=1}^{K-1}\left( \frac{1}{(a_{k+1})^{\alpha }}-\frac{1}{%
(a_{k})^{\alpha }}\right) \text{ .}
\end{equation*}
\end{proof}

\subsection{\label{6.4}proposition \ref{4c}}

We use the shorthand notation 
\begin{equation*}
\left( U(t);V(t)\right) =\left( \frac{L(t)-\left( X_{1}(t)+\cdots
+X_{n+1}(t)\right) }{X_{n+1}(t)}\text{ ; }\frac{X_{1}(t)}{X_{n+1}(t)},\cdots
,\frac{X_{n}(t)}{X_{n+1}(t)}\right) \text{ .}
\end{equation*}

\begin{proof}
(Recall that $\lambda (dx)=\alpha x^{-(1+\alpha )}dx$ and $\Lambda
(x)=1/x^{\alpha }$)\bigskip

\textbf{Step 1}

Due to the `Poissonian-superposition' structure of the OLP $L$, given the
event $\{X_{n+1}(t)=x\}$ the vector $\left( U(t);V(t)\right) $ is equal, in
law, to the vector: 
\begin{equation*}
\left( \frac{S^{x}(t)}{x}\text{ ; }\frac{J_{(1)}^{x}}{x},\cdots ,\frac{%
J_{(n)}^{x}}{x}\right)
\end{equation*}
where:

\textbf{(i)} $S^{x}(t)$ is the value, at time $t$, of a OLP with L\'{e}vy
measure 
\begin{equation}
\lambda ^{x}(dy)=\left\{ 
\begin{array}{ccc}
\lambda (dy) &  & 0<y<x \\ 
&  &  \\ 
0 &  & y>x
\end{array}
\right. \text{ .}  \label{6.41}
\end{equation}

\textbf{(ii)} $J_{(1)}^{x},\cdots ,J_{(n)}^{x}$ are the order statistics of $%
J_{1}^{x},\cdots ,J_{n}^{x}$ which, in turn, are IID random variables
distributed according to the probability law (supported on the half line $%
(x,\infty )$): 
\begin{equation}
\text{\textbf{P}}\left( J^{x}\in dy\right) =\left\{ 
\begin{array}{ccc}
0 &  & 0<y<x \\ 
&  &  \\ 
\lambda (dy)/\Lambda (x) &  & y>x
\end{array}
\right. \text{ .}  \label{6.42}
\end{equation}

\textbf{(iii) }$S^{x}(t)$ and $J_{1}^{x},\cdots ,J_{n}^{x}$ are mutually
independent.

\bigskip

Hence, $\forall \omega \geq 0$ and $\forall \theta =(\theta _{1},\cdots
,\theta _{n})\geq 0$ we have 
\begin{equation*}
\text{\textbf{E}}\left[ \exp \left\{ -\omega U(t)-\theta V(t)\right\} \text{ 
\TEXTsymbol{\vert} }X_{n}(t)=x\right] =\text{\textbf{E}}\left[ \exp \left\{ 
\frac{-\omega }{x}S^{x}(t)\right\} \right] \cdot \text{\textbf{E}}\left[
\exp \left\{ \frac{-\theta }{x}\overrightarrow{J^{x}}\right\} \right] \text{
,}
\end{equation*}
where $\overrightarrow{J^{x}}=(J_{(1)}^{x},\cdots ,J_{(n)}^{x})$.

\bigskip

\textbf{Step 2}

Since $S^{x}(t)$ is the value, at time $t$, of a OLP with L\'{e}vy measure (%
\ref{6.41}) we have 
\begin{equation*}
\text{\textbf{E}}\left[ \exp \left\{ \frac{-\omega }{x}S^{x}(t)\right\} %
\right] =\exp \left\{ -\Psi ^{x}\left( \frac{\omega }{x}\right) t\right\} 
\text{ ,}
\end{equation*}
where 
\begin{equation*}
\left. 
\begin{array}{l}
\Psi ^{x}\left( \frac{\omega }{x}\right) \\ 
\\ 
=\int_{0}^{x}\left( 1-\exp \left\{ \frac{-\omega }{x}y\right\} \right)
\lambda (dy) \\ 
\\ 
=\frac{\alpha }{x^{\alpha }}\int_{0}^{1}\frac{1-\exp \{-\omega u\}}{%
u^{1+\alpha }}du \\ 
\\ 
=\Lambda (x)G_{\alpha }(\omega )\text{ .}
\end{array}
\right.
\end{equation*}

On the other hand, using (\ref{6.42}), $\forall y>1$ we have: 
\begin{equation*}
\text{\textbf{P}}\left( \frac{J^{x}}{x}>y\right) =\frac{\Lambda (xy)}{%
\Lambda (x)}=\frac{1}{y^{\alpha }}\text{ .}
\end{equation*}

\bigskip

Hence, combining the above together, we conclude that 
\begin{equation}
\text{\textbf{E}}\left[ \exp \left\{ -\omega U(t)-\theta V(t)\right\} \text{ 
\TEXTsymbol{\vert} }X_{n}(t)=x\right] =\exp \left\{ -\Lambda (x)G_{\alpha
}(\omega )t\right\} \cdot L(\theta )\text{ ,}  \label{6.43}
\end{equation}
where $L(\theta )$ is the Laplace transform (at the point $\theta =(\theta
_{1},\cdots ,\theta _{n})$) of the \emph{order statistics} of $%
\{Y_{1},\cdots ,Y_{n}\}$ which are IID Pareto$(\alpha )$-distributed random
variables: \textbf{P}$\left( Y>y\right) =1/y^{\alpha }$, $y\geq 1$.

\bigskip

\textbf{Step 3}

Using (\ref{2.12}), the probability density function of $X_{n+1}(t)$ is
given by 
\begin{equation}
\left. 
\begin{array}{l}
\text{\textbf{P}}\left( X_{n+1}(t)\in dx\right) \\ 
\\ 
=P_{1}\left( t\lambda (dx)\right) \cdot P_{n}\left( t\int_{x}^{\infty
}\lambda (ds)\right) \\ 
\\ 
=\left( t\lambda (dx)\right) \cdot \frac{\left( t\Lambda (x)\right) ^{n}}{n!}%
\exp \left\{ -t\Lambda (x)\right\} \\ 
\\ 
=\frac{t^{n+1}}{n!}\exp \left\{ -t\Lambda (x)\right\} \Lambda (x)^{n}\lambda
(dx)\text{ .}
\end{array}
\right.  \label{6.44}
\end{equation}

Combining (\ref{6.43}) and (\ref{6.44}) together we hence obtain 
\begin{equation*}
\left. 
\begin{array}{l}
\text{\textbf{E}}\left[ \exp \left\{ -\omega U(t)-\theta V(t)\right\} \right]
\\ 
\\ 
=\int_{0}^{\infty }\text{\textbf{E}}\left[ \exp \left\{ -\omega U(t)-\theta
V(t)\right\} \text{ \TEXTsymbol{\vert} }X_{n}(t)=x\right] \text{\textbf{P}}%
\left( X_{n}(t)\in dx\right) \\ 
\\ 
=I\cdot L(\theta )\text{ ,}
\end{array}
\right.
\end{equation*}
where 
\begin{equation*}
I=\frac{t^{n+1}}{n!}\int_{0}^{\infty }\exp \left\{ -\Lambda (x)G_{\alpha
}(\omega )t\right\} \cdot \exp \left\{ -t\Lambda (x)\right\} \Lambda
(x)^{n}\lambda (dx)\text{ .}
\end{equation*}

However, using the change of variable $s=\Lambda (x)$ gives 
\begin{equation*}
I=\frac{t^{n+1}}{n!}\int_{0}^{\infty }\exp \left\{ -t(1+G_{\alpha }(\omega
))\cdot s\right\} s^{n}ds=\frac{1}{\left( 1+G_{\alpha }(\omega )\right)
^{n+1}}\text{ ,}
\end{equation*}
and hence we can conclude that 
\begin{equation}
\text{\textbf{E}}\left[ \exp \left\{ -\omega U(t)-\theta V(t)\right\} \right]
=\left( \frac{1}{1+G_{\alpha }(\omega )}\right) ^{n+1}\cdot L(\theta )\text{
.}  \label{6.45}
\end{equation}

Equation (\ref{6.45}) implies that $\left( U(t);V(t)\right) =\left( \xi
_{1}+\cdots +\xi _{n+1}\text{ ; }Y_{(1)},\cdots ,Y_{(n)}\right) $ where:

\qquad \textbf{(i)} $\xi _{1},\cdots ,\xi _{n+1},Y_{1},\cdots ,Y_{n}$ are
independent random variables;

\qquad \textbf{(ii)} $\xi _{1},\cdots ,\xi _{n+1}$\ are $\mathcal{G}_{\alpha
}$-distributed;

\qquad \textbf{(iii)} $Y_{(1)},\cdots ,Y_{(n)}$ are the order statistics of $%
Y_{1},\cdots ,Y_{n}$ which are Pareto$(\alpha )$-distributed: \textbf{P}$%
\left( Y>y\right) =1/y^{\alpha }$, $y\geq 1$.
\end{proof}


\begin{thebibliography}{99}
\bibitem{L25}  P. L\'{e}vy, Calcul des Probabilit\'{e}s, Gauthier-Villars,
1925.

\bibitem{L54}  P. L\'{e}vy, Th\'{e}orie de l'addition des variables
Al\'{e}atoires, Gauthier-Villars, 1954.

\bibitem{L65}  P. L\'{e}vy, Processus Stochastiques et mouvement Brownien,
Gauthier-Villars, 1965.

\bibitem{GK}  B.V. Gnedenko \& A.N. Kolmogorov, Limit distributions for Sums
of Independent Random Variables, Addison-Wesley, 1954.

\bibitem{IL}  I.A. Ibragimov \& Yu.V. Linnik, Independent and Stationary
Sequences of Random Variables, Walterss-Noordhoff, 1971.

\bibitem{Fel}  W. Feller, An Introduction to Probability Theory and Its
Applications, Vol. II, 2$^{\text{nd}}$ edition, Wiley, 1971.

\bibitem{ZOL}  V.M. Zolotarev, One-Dimensional Stable Distributions, AMS,
1986.

\bibitem{ST}  G. Samrodintsky \& M.S. Taqqu, Stable non-Gaussian Random
Processes, CRC Press, 1994.

\bibitem{Ber}  J. Bertoin, L\'{e}vy Processes, Cambridge University Press,
1996.

\bibitem{Sat}  K. Sato, L\'{e}vy Processes and Infinitely Divisible
Distributions, Cambridge University Press, 1999.

\bibitem{UZ}  V.V. Uchaikin \& V.M. Zolotarev, Chance and Stability, Stable
Distributions and Their Applications, V.S.P. Intl. Science, 1999.

\bibitem{BMR}  O.E. Barndorff-Nielsen, T. Mikosch, \& S. Resnic (Eds.),
L\'{e}vy Processes, Birkhauser, 2001.

\bibitem{BG}  J.P. Bouchaud \& A. Georges, Phys. Rep. 195 (1990) 12.

\bibitem{SZK}  M.F. Shlesinger, G.M. Zaslavsky, \& J. Klafter, Nature 363
(1993) 31.

\bibitem{SZF}  M.F. Shlesinger, G.M. Zaslavsky, \& U. Frisch (Eds.),
L\'{e}vy Flights and Related Topics, Springer, 1995.

\bibitem{KSZ}  J. Klafter, M.F. Shlesinger, \& G. Zumofen, Phys. Today 49
(2) (1996) 33.

\bibitem{Hav}  G.M. Viswanathan et al., Nature 401 (1999) 911; G.M.
Viswanathan et al., Physica A 282 (2000) 1; G.M. Viswanathan et al., Physica
A 295 (2001) 85.

\bibitem{Bor}  L. von Bortkiewicz, Sitzungsber. Berlin Math. Ges. 21 (1922)
3.

\bibitem{Fre}  M. Fr\'{e}chet, Ann. Soc. Polon. Math. Cracovie, 6 (1927) 93.

\bibitem{FT}  R.A. Fisher \& L.H.C. Tippett, Procs. Cambridge Philos. Soc.,
24 (1928) 180.

\bibitem{Mis}  R. von Mises, Rev. Math. Union Interbalk, 1 (1936) 141
(reprinted in: Selected Papers of von Mises II, pp. 271-294, Amer. Math.
Soc., 1954).

\bibitem{Gum}  E.J.\ Gumbel, Statistics of Extremes, Columbia University
Press, 1958.

\bibitem{Gne}  B. Gnedenko, Ann. Math. 44 (1943) 423 (translated and
reprinted in: Breakthroughs in Statistics I, edited by S. Kotz \& N.L.
Johnson, pp. 195-225, Springer, 1992).

\bibitem{EKM}  P. Embrechts, C. Kluppelberg \& T. Mikosch, Modelling
extremal events for insurance and finance, Springer 1997.

\bibitem{KN}  S. Kotz \& S. Nadarajah, Extreme value distributions, Imperial
College Press, 2000.

\bibitem{TR}  M. Thomas \& R.D. Reiss, Statistical analysis of extreme
values, Birkhauser, 2001.

\bibitem{MANA}  A.A. Moreira, J.S. Andrade, \& L.A. Nunes Amaral, Phys. Rev.
Lett. 89 (26) (2002) 268703-1.
\end{thebibliography}
\end{document}